\documentclass[two column,jgrga]{agutex}

\usepackage[dvips]{graphicx}

\authorrunninghead{DALLA ET AL.}

\titlerunninghead{DRIFTS IN PARKER SPIRAL}

\begin{document}

%
%

\title{Solar Energetic Particle drifts in the Parker spiral}
%
%

%
%



\authors{S.~Dalla,\altaffilmark{1}
M.S.~Marsh,\altaffilmark{1} J.~Kelly,\altaffilmark{1}
and T.~Laitinen\altaffilmark{1}}

\altaffiltext{1}{Jeremiah Horrocks Institute,
University of Central Lancashire, Preston, UK. Email: sdalla@uclan.ac.uk}





%
%


\begin{abstract}
Drifts in the Parker spiral interplanetary magnetic field are known to be an important component in the propagation of galactic cosmic rays, while they are thought to be negligible for Solar Energetic Particles (SEPs). As a result they have so far been ignored in SEP propagation modelling and data analysis. We examine drift velocities in the Parker spiral within single particle first-order adiabatic theory, in a local coordinate system with an axis parallel to the magnetic field. We show that, in the presence of scattering in interplanetary space, protons at the high end of the SEP energy range experience significant gradient and curvature drift. In the scatter-free case, drift due to magnetic field curvature is present. The magnitude of drift velocity increases by more than an order of magnitude at high heliographic latitudes compared to near the ecliptic; it has a strong dependence on radial distance $r$ from the Sun, reaching a maximum at $r$$\sim$1 AU at low heliolatitudes and $r$$\sim$10 AU at high heliolatitudes. Due to the mass over charge dependence of drift velocities, the effect of drift for partially ionised SEP heavy ions is stronger than for protons. Drift is therefore likely to be a considerable source of cross field transport for high energy SEPs.    
\end{abstract}

%
%

%

\begin{article}

%
%

\section{Introduction}

The large-scale structure of the interplanetary magnetic field (IMF) can be approximated as a Parker spiral  \citep{Par1958}. 
Although it is known that this basic structure is altered by turbulence and by a variety of other effects, including the tilt of the magnetic axis with respect to the rotation axis, differential rotation and transients, a variety of spacecraft measurements have confirmed that the large-scale spatial variation of the IMF broadly follows Parker's model (e.g.~\cite{For1996}).

Energetic charged particles are guided by the IMF in their propagation, and scattered by its turbulence. They experience guiding centre drifts due to the presence of magnetic field inhomogeneity, curvature, and the solar wind electric field. Drift effects can be treated either by means of single particle first-order guiding centre theory \citep{Nor1963,Ros1970} or within kinetic theory (for a discussion of these approaches and their connections see \citet{Bur1985}).

The importance of drifts in the propagation of Galactic Cosmic Rays (GCRs) is well established and is usually modelled by means of the formulation by \cite{Jok1977} within the framework of the Parker transport equation. For Solar Energetic Particles (SEPs), characterised by energies lower than GCRs, drift effects are considered to be unimportant. In many instances SEPs are assumed to be tied to magnetic field lines. This assumption is intrinsic within the majority of SEP propagation models, starting from the formulation of the focussed transport equation \citep{Roe1969}, and up to modern efforts based on similar approaches [e.g.~\citet{Ruf1995, Lar1998, Luh2007}].
In the latter works, the assumption that particles are tied to magnetic field lines is used to reduce the number of spatial variables required in the modelling to one, the distance travelled along the field line, thus simplifying the problem of solving the corresponding equations.

Single particle first-order guiding centre drifts for SEPs in a Parker spiral configuration were considered in two early studies which concluded that they play a negligible role. The first of these studies \citep{Bur1968} was based on an analytical calculation of drift velocities, while the second used  numerical integration \citep{Win1968}. Both approaches assumed that particles propagate scatter-free, hence experiencing strong focussing in the IMF, so that in practice a zero particle pitch-angle was used in the calculation of drift velocities. In addition, drift velocities were only calculated in the heliographic equatorial plane.

However, spacecraft measurements in interplanetary space clearly show that SEPs can have large pitch-angles at locations far away from the Sun. Even during the early phase of SEP events, a fraction of particles have pitch-angles close to 90$^{\circ}$ and this fraction increases greatly during an event, with anisotropies in most cases becoming negligibly small during the peak and decay phases.  

In this paper, we investigate single particle first-order drifts in the Parker spiral without assuming a zero pitch-angle. On the contrary, we allow for the possibility that particles may be characterised by a broad range of pitch-angles while propagating in the interplanetary medium. We analyse the magnitude of drifts as a function of location in the heliosphere.
We show that magnitudes of drift velocities are significant for energetic particles at the high energy end of the SEP range, and are a non-negligible cause of transport across the field.  

In Section \ref{sec.magelec} we derive analytical relativistic expressions for single particle first-order drifts in a local reference frame with an axis along the Parker spiral. In Section \ref{sec.drift_mag_sep} magnitudes of drift velocities and their spatial variation are analysed for particles in the SEP energy range. Section \ref{sec.discussion} presents a discussion of the results and conclusions.

\section{Single-particle drift velocities in the Parker spiral}\label{sec.magelec}

\subsection{Unipolar Parker field} \label{sec.unipolar}

The Parker spiral interplanetary magnetic field is given by \citet{Par1958}:

\begin{eqnarray}
B_r &=& B_0 \,\frac{r_0^2}{r^2} \label{eq.br}\\
B_{\theta} &=& 0  \\
B_{\phi} &=& - \frac{B_0  \, r_0^2 \, \Omega}{v_{sw}} \, \frac{\sin{\theta}}{r}  \label{eq.bphi}
\end{eqnarray}
where ($r,\theta,\phi$) are heliocentric spherical coordinates 
with $r$ the radial distance, $\theta$ the colatitude and $\phi$ the longitude. Here $B_0$ is the magnitude of the 
magnetic field at a reference distance $r_0$, 
$\Omega$ the solar rotation rate (taken as constant) and $v_{sw}$ the solar wind speed. A solar wind flow that is radial, uniform and time independent is assumed.

Eqs.~(\ref{eq.br})--(\ref{eq.bphi}) describe a unipolar field pointing away from the Sun. In reality, the IMF is characterised by at least two domains of opposite polarity, separated by a current sheet. In this Section we analyse drift velocities in the simplified unipolar field given by Eqs.~(\ref{eq.br})--(\ref{eq.bphi}), while the effect of the presence of two polarities will be described in Section \ref{sec.dipolar}.

Due to the motion of the solar wind, in the inertial (non-rotating) reference frame an electric field $\mathbf{E} = -  \mathbf{v}_{sw} /c \times \mathbf{B}$ is present, which, using Eqs.~(\ref{eq.br})--(\ref{eq.bphi}), takes the form:
\begin{eqnarray}
E_r &=& 0 \label{eq.er}\\
E_{\theta} &=& - \frac{\Omega  \, B_0 r_0^2}{c} \,  \, \frac{\sin{\theta}}{r}   \label{eq.etheta}  \\
E_{\phi} &=& 0 \label{eq.ephi}
\end{eqnarray}
where $c$ is the speed of light.

\cite{Bur1968} [from hereon indicated as BH1968] calculated
particle drift velocities in the magnetic and electric fields of Eqs.~(\ref{eq.br})--(\ref{eq.ephi}) analytically in spherical coordinates, for a single particle of nonrelativistic speed and within the assumptions of standard first-order adiabatic theory. The drift velocities consist of an electric field drift $\mathbf{v}_{E}$, a grad-$B$ drift $\mathbf{v}_{\nabla B}$, a curvature drift $\mathbf{v}_{c}$ and a so-called polarisation drift $\mathbf{v}_{p}$ [given by Eqs (10), (12), (14) and (15) of BH1968 respectively].

We introduce a local coordinate system $(\mathbf{e}_{l},\mathbf{e}_{\phi'},\mathbf{e}_{\theta'})$ with an axis parallel to the Parker spiral \citep{Kel2012} and 
calculate the components of the drift velocities in this system. Compared with spherical coordinates, this choice of coordinate system has the advantage that the analytical expressions take a simpler form, as they have at most two nonzero components, both perpendicular to the field.
The local coordinate system has an axis $\mathbf{e}_{l}$ along the direction of the Parker spiral and
pointing outwards, another axis in the direction of $\mathbf{e}_{\theta'}$=$-$$\mathbf{e}_{\theta}$ with $\mathbf{e}_{\theta}$ the standard spherical coordinate system unit vector and an axis $\mathbf{e}_{\phi'}$  completing the right-handed orthogonal system.

In the local Parker system, the electric field drift velocity is given by:
\begin{eqnarray}
v_{El} &=& 0 \label{eq.parkerelec1} \\
v_{E\phi'} &=&  \frac{v_{sw} \, r}{(r^2+a^2)^{1/2}} \\
v_{E\theta'} &=& 0  \label{eq.parkerelec3} 
\end{eqnarray}
where $a$ is a function of colatitude $\theta$ and is defined as:
\begin{equation}
a =  \frac{v_{sw}}{\Omega \, \sin{\theta}}. 
\end{equation}
The electric field drift is always in the $\mathbf{e}_{\phi'}$ direction and is independent of particle properties such as speed, charge and mass. This drift describes the corotation of a particle with the IMF magnetic field lines as the Sun rotates, i.e.~it is a corotation drift. Near the Sun, it moves particles in the direction of solar rotation by 14.3$^{\circ}$ per day.

The grad-$B$ drift velocity has the expression:
\begin{eqnarray}
v_{\nabla B\, l} &=& 0 \label{eq.parkergradb1} \\
v_{\nabla B\, \phi'} &=&  \frac{\mu  c}{q} \frac{1}{r^2+a^2} \, \, r  \cot{\theta} \label{eq.parkergradb2}\\
v_{\nabla B\, \theta'} &=&  - \frac{\mu  c}{q} \frac{1}{(r^2+a^2)^{3/2}} \,\, (r^2+2a^2) \label{eq.parkergradb3} 
\end{eqnarray}
where $\mu$ is the particle's magnetic moment and $q$ its charge.
The grad-$B$ drift depends on the particle species and on velocity. The direction of the grad-$B$ drift is opposite
for electrons and ions. 
In the nonrelativistic approximation, the magnetic moment is given by:
\begin{equation}\label{eq.magnmom}
\mu= \frac{m v_{\perp}^2}{2  B}
\end{equation}
where  $v_{\perp}$ is the component of a particle's velocity in a plane perpendicular to the magnetic field and $B$=$|\mathbf{B}|$. 
Given its direct proportionality to $\mu$,  $\mathbf{v}_{\nabla B}$ 
is largest for particles with pitch-angle $\alpha$=90$^{\circ}$ and equal to zero for field aligned (strongly focussed) particles. 

The curvature drift has the expression:
\begin{eqnarray}
v_{c l} &=& 0 \\
v_{c \phi'} &=&  - \frac{m c}{qB} \, v_{\|}^2    \frac{1}{r^2+a^2} \,\, \, r\,\cot{\theta}\\
v_{c \theta'} &=&  - \frac{m c}{qB} \, v_{\|}^2   \frac{1}{(r^2+a^2)^{3/2}} \,\, (r^2+2a^2)
\end{eqnarray}
where $v_{\|}$ is the component of the particle's velocity parallel to the magnetic field. The curvature drift has a very similar expression to the grad-$B$ drift, but because of its dependence on  $v_{\|}$, it is largest in magnitude for particles with $\alpha$=0$^{\circ}$ (parallel motion) and  $\alpha$=180$^{\circ}$ (anti-parallel motion).
The polarisation drift is given by:
\begin{eqnarray}
v_{ pl} &=& 0 \\
v_{ p \phi'} &=&  \frac{m c}{qB}   \frac{v_{sw} v_{\|} }{(r^2+a^2)^{3/2}} \,\, a r \cot{\theta} \\
v_{ p \theta'} &=&  \frac{m c}{qB}  \frac{v_{sw} v_{\|}}{(r^2+a^2)^{2}}\,\, a^3  \label{eq.polaris3}
\end{eqnarray}
The grad-$B$, curvature and polarisation drift velocities increase with particle energy.

A generalisation of Eqs.~(\ref{eq.parkergradb1})--(\ref{eq.polaris3}) to the case of relativistic particles can be obtained by letting the particle mass $m$ take its relativistic form $m$=$m_0\, \gamma$ where $m_0$ is the particle's rest mass and $\gamma$ is the relativistic Lorentz factor (see Eq.(5.64) of \cite{Ros1970}). We will be adopting relativistic expressions in the following.

In the limit of scatter-free propagation, the magnetic moment $\mu$ is a constant of motion and this results in strong focussing due to the decrease of the magnetic field magnitude with $r$. If scattering is present, $\mu$ is no longer a constant since scattering events give rise to changes in pitch-angle, and the evolution of a particle's pitch-angle is not predetermined, as it would be in the scatter-free case.  
BH1968 and \cite{Win1968} assumed a constant $\mu$ when calculating grad-$B$ drift velocities, so that their results only apply to particles propagating scatter-free.
When $\mu$ is treated as constant, Eqs.~(\ref{eq.parkergradb2})--(\ref{eq.parkergradb3}) give the grad-$B$ drift velocity of a particle as it travels to different $r$-values while focussing. 

In this paper we allow for the possibility of scattering taking place and consequently we do not set $\mu$ to a constant in the expressions for the grad-$B$ drift.
Instead, we expand  $\mu$ in  Eqs.~(\ref{eq.parkergradb2})--(\ref{eq.parkergradb3}) using Eq.(\ref{eq.magnmom}).  

Using Eq.(\ref{eq.magnmom}) and the expression for the magnitude of $\mathbf{B}$ in the Parker spiral:
\begin{equation}
B= \frac{B_0 r_0^2}{r^2 a} \, (r^2+a^2)^{1/2} 
\end{equation}
Eqs.(\ref{eq.parkergradb1})--(\ref{eq.parkergradb3}) become, in relativistic form:
\begin{eqnarray}
v_{\nabla B\, l} &=& 0  \label{eq.gradbdrift_l} \\
v_{\nabla B\, \phi'} &=& \frac{1}{2} \frac{m_0 \gamma \, c}{q} v_{\perp}^2 \, g(r,\theta)\\
v_{\nabla B\, \theta'} &=&  - \frac{1}{2} \frac{m_0 \gamma \, c}{q} v_{\perp}^2 \, f(r,\theta) \label{eq.gradbdrift_thetapr}
\end{eqnarray}
where:
\begin{eqnarray}
g(r,\theta) &=&  \frac{a}{B_0 r_0^2} \, \frac{x^3 \cot{\theta}}{(x^2+1)^{3/2}}  \label{eq.g_function}\\
f(r,\theta) &=& \frac{a}{B_0 r_0^2} \, \frac{x^2(x^2+2)}{(x^2+1)^2} \label{eq.f_function}
\end{eqnarray}
and $x$=$x(r,\theta)$=$r/a(\theta)$. Here $g(r,\theta)$ describes the spatial variation of the $\phi'$ components of the grad-$B$ drift, and $f(r,\theta)$ that of the $\theta'$ components.

\begin{figure}
\noindent\includegraphics[width=20pc]{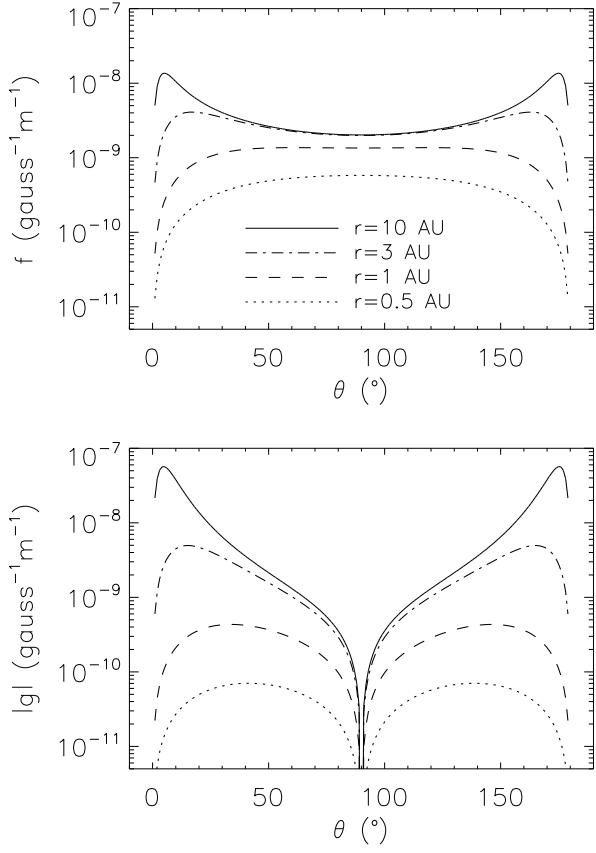}
\caption{$f$ ({\it top panel}) and $|g|$ ({\it bottom panel}) versus colatitude $\theta$ for several values of $r$. Note that $g$$>$0 for $\theta$$\in$[0$^{\circ}$, 90$^{\circ}$] and  $g$$<$0 for $\theta$$\in$[90$^{\circ}$, 180$^{\circ}$].}
  \label{fig.fabsgvstheta}
\end{figure}

\begin{figure*}
\noindent\includegraphics[width=39pc]{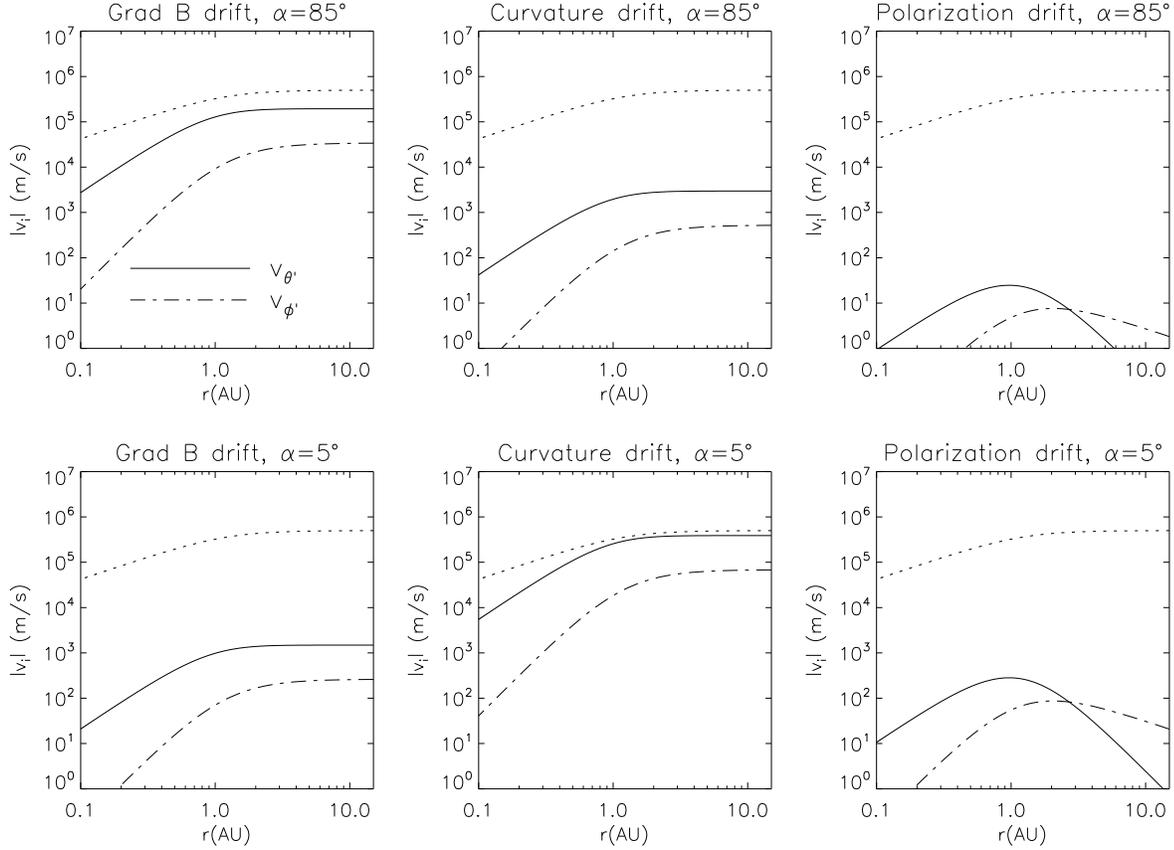}
\caption{Absolute values of drift velocity components in the local Parker coordinate system for a 100 MeV proton at latitude $\delta$=10$^{\circ}$, as a function of $r$. Panels show, from left to right: grad-$B$, curvature and polarisation drift components.  The solid line gives the $\theta'$ component and the dash dotted line the $\phi'$ one. In all panels the dotted line is the corotation speed  $v_{E \phi'}$. {\it Top row:} Pitch-angle $\alpha$=85$^{\circ}$; {\it bottom row:} $\alpha$=5$^{\circ}$.}
\label{fig.driftscolat80}
\end{figure*}

Similarly we can express the curvature drift as:
\begin{eqnarray}
v_{c l} &=& 0 \label{eq.curvdrift_l}\\
v_{c \phi'} &=&  - \frac{m_0 \gamma \, c}{q} \, v_{\|}^2   \, g(r,\theta)  \\
v_{c \theta'} &=&  - \frac{m_0 \gamma \, c}{q} \, v_{\|}^2 \, f(r,\theta) \label{eq.curvdrift_thetapr}
\end{eqnarray}
and the polarisation drift as:
\begin{eqnarray}
v_{ pl} &=& 0 \\
v_{ p \phi'} &=&  \frac{m_0 \gamma \, c}{qB}   \frac{v_{sw} v_{\|} }{(r^2+a^2)^{3/2}} \,\, a r \cot{\theta} \\
v_{ p \theta'} &=&  \frac{m_0 \gamma \, c}{qB}  \frac{v_{sw} v_{\|}}{(r^2+a^2)^{2}}\,\, a^3  \label{eq.polaris3_rel}
\end{eqnarray}


Eqs.(\ref{eq.parkerelec1})--(\ref{eq.parkerelec3}) and (\ref{eq.gradbdrift_l})--(\ref{eq.polaris3_rel}) constitute the full set of first-order drift velocities for a relativistic particle of arbitrary pitch-angle in the unipolar Parker spiral, expressed in the local Parker coordinate system. This set of equations is equally applicable to SEPs and GCRs.

Due to the dependence on $\cot{\theta}$, the function $g$ and consequently the  $\phi'$ components of the grad-$B$  and curvature drift velocities are zero at the heliographic equator (where $\theta$=90$^{\circ}$). Also, $g$$>$0 when $\theta$$\in$[0$^{\circ}$, 90$^{\circ}$] and  $g$$<$0 when $\theta$$\in$[90$^{\circ}$, 180$^{\circ}$], while $f$ is always positive.
Figure \ref{fig.fabsgvstheta} shows the variation of $f$ and $|g|$ with $\theta$, at four values of $r$.
The function $f$, influencing the $\theta'$ components of drift velocities, has a fairly slow variation with $\theta$, so that the magnitude of drift in $\theta'$ does not depend strongly on colatitude.
On the other hand 
$g$ has a strong $\theta$-dependence, so that the drift in $\phi'$ is zero near the heliographic equator but increases greatly at high heliographic latitudes.

For a unipolar Parker field pointing outwards (as in  Eqs.(\ref{eq.br})--(\ref{eq.bphi}))
the directions of the grad-$B$ and curvature drifts for a positively charged ion are as follows.
$v_{\nabla B \theta'}$ is always negative, therefore in regions not far from the heliographic equatorial plane (where the $\mathbf{e}_{\theta'}$ direction points approximately upwards)  the grad-$B$ drift tends to push particle downwards (both above and below the plane). Similarly the curvature component $v_{c \theta'}$ is always negative, hence in the same direction as $v_{\nabla B \theta'}$.
For $\theta$$\in$[0$^{\circ}$, 90$^{\circ}$] (i.e.~$g$$>$0) $v_{\nabla B \phi'}$ is positive (i.e.~roughly in the direction of to corotation near the equatorial plane), while $v_{c \phi'}$ is negative (i.e.~roughly opposite corotation near the equatorial plane). The signs of the $\phi'$ components of the drifts reverse for $\theta$$\in$[90$^{\circ}$, 180$^{\circ}$]. 

The directions of grad-$B$ and curvature drifts reverse for an electron due to the dependence on $q$.

\subsection{Dipolar Parker field} \label{sec.dipolar}

Unlike in the unipolar magnetic field of Eqs.~(\ref{eq.br})--(\ref{eq.bphi}), in the solar magnetic field, and consequently the IMF, two polarities are present.
At solar minimum, the IMF's structure can be broadly described as two hemispheres of opposite polarities, separated by a current sheet, called the Heliospheric Current Sheet (HCS).
At solar maximum, a more complex structure consisting of dipolar plus quadrupolar contributions exists. Since the tilt of the magnetic dipole increases as the cycle progresses from minimum to maximum, the tilt angle of the HCS also increases.

In this Section, we extend the discussion of energetic particle drifts of Section \ref{sec.unipolar} to a simplified dipolar IMF, where the HCS coincides with the heliographic equatorial plane and the magnetic field points outwards everywhere in the northern hemisphere and inwards everywhere in the southern hemisphere. This configuration is usually described as $A^+$ in studies of GCRs, and the one with opposite polarities as $A^-$.

In an $A^+$ configuration, drifts for particles in the northern hemisphere are the same as described in Section \ref{sec.unipolar}. To derive drift directions in the southern hemisphere, we consider how these change when $-\mathbf{B}$ is substituted in place of $\mathbf{B}$ in the expressions for drift velocities. As expected, the electric field drift (corotation drift) remains in the same direction because the electric field direction also reverses. The grad-$B$ drift velocity takes opposite signs to those in Eqs.(\ref{eq.gradbdrift_l})--(\ref{eq.gradbdrift_thetapr}) and so does the curvature drift velocity with respect to Eqs.~(\ref{eq.curvdrift_l})--(\ref{eq.curvdrift_thetapr}). Hence for a positively charged particle in the southern hemisphere ($\theta$$\in$[90$^{\circ}$, 180$^{\circ}$]) the grad-$B$ and curvature drifts will be positive along the $\mathbf{e}_{\theta'}$ direction, thus pushing particles upwards at colatitudes not far from the heliographic equator. In the $\mathbf{e}_{\phi'}$ directions drifts will be opposite compared to those discussed in Section \ref{sec.unipolar}.
Therefore, as expected, the directions of drifts from single-particle first-order guiding centre theory for an $A^+$ configuration give rise to the well-known GCR pattern, where positively charged particles drift towards the HCS \citep{Jok1977}. It should be noted that the standard GCR drift expressions are derived for an isotropic particle distribution and do not contain a dependence on pitch-angle.

In an $A^-$ configuration, the grad-$B$ and curvature drift patterns are reversed, with positively charged particles drifting from low to high heliolatitudes, i.e.~away from the HCS.

If we define a drift velocity $\mathbf{v}_d$ as the sum of the grad-$B$ and curvature drifts:
\begin{equation}
\mathbf{v}_d = \mathbf{v}_{\nabla B} + \mathbf{v}_{c}
\end{equation}
using Eqs.~(\ref{eq.gradbdrift_l})--(\ref{eq.gradbdrift_thetapr}), Eqs.~(\ref{eq.curvdrift_l})--(\ref{eq.curvdrift_thetapr}) and keeping into account the drift directions in the $A^+$ and $A^-$ configurations, we can write the following summary expression for $\mathbf{v}_d$, valid in a simple dipolar field with current sheet coinciding with the heliographic equatorial plane:

\begin{eqnarray} 
v_{d l} &=& 0 \label{eq.vdl}\\
v_{d \phi'} & = & A \:\, \mbox{sgn}\left(\frac{\pi}{2}-\theta \right) \, \frac{m_0 \gamma \,  c}{q} \left( \frac{1}{2} \,v_{\perp}^2 - v_{\|}^2 \right) \, g(r,\theta) \label{eq.vdphipr}\\ 
v_{d \theta'} & = & - A \:\, \mbox{sgn}\left(\frac{\pi}{2}-\theta \right) \, \frac{m_0 \gamma \,  c}{q} \left( \frac{1}{2} \,v_{\perp}^2 + v_{\|}^2 \right) \, f(r,\theta) \label{eq.vdthetapr}
\end{eqnarray}
where $A$=1 during an  $A^+$ cycle, $A$=$-$1 during an $A^-$ cycle, $\mbox{sgn}$ is the sign function and the colatitude $\theta$ is expressed in radians.

When particles get within two gyroradii of the heliospheric current sheet, their trajectories are no longer described by adiabatic theory and a separate analysis is required.
It can be shown that the effect of a neutral sheet on charged particle motion is to give rise to a drift along the sheet itself and perpendicular to the direction of the magnetic field \citep{Bur1985}. Current sheet drift is important in the propagation of GCRs and a number of different approaches for describing it have been proposed (see e.g.~\citet{Bur1989} and references therein). Its importance for SEP propagation will need to be established in future work.

\section{Drift velocities for SEPs} \label{sec.drift_mag_sep}

\subsection{Protons}

As far as we are aware, at present drifts are not taken into account in most models of SEP propagation and are thought to be unimportant in the SEP energy range.

Figure \ref{fig.driftscolat80} shows the absolute value of drift velocity components for a 100 MeV proton at heliographic latitude $\delta$=10$^{\circ}$ (i.e.~colatitude $\theta$=80$^{\circ}$). In each row, the panels give, from left to right,  $\mathbf{v}_{\nabla B}$, $\mathbf{v}_{c}$ and $\mathbf{v}_{p}$ and the dotted line in all panels represents the corotation velocity $v_{E \phi'}$. The top row is for a particle of pitch-angle $\alpha$=85$^{\circ}$ and  the bottom row for one with $\alpha$=5$^{\circ}$. The relativistic factor for a 100 MeV proton is $\gamma$=1.1. We use $v_{sw}=$500 km s$^{-1}$, $\Omega$=2.86$\times$10$^{-6}$ rad s$^{-1}$, $B_0$=1.78 gauss and $r_0$=1 $r_s$ with $r_s$ the solar radius ($r_s$=6.96$\times$10$^{8}$ m); this choice of $B_0$ and $r_0$ ensures that the magnetic field magnitude at 1 AU is 5 nT.

Comparing the top and bottom rows in Figure \ref{fig.driftscolat80} it is apparent that for a particle with $\alpha$$\sim$90$^{\circ}$  the grad-$B$ drift is dominant, while for $\alpha$$\sim$0$^{\circ}$ it is the curvature drift that dominates. The polarisation drift is much smaller than the sum of grad-$B$ and curvature drifts for the case of Figure \ref{fig.driftscolat80} and all other cases we considered, and will not be further discussed. The magnitudes of the grad-$B$ and curvature drifts can reach values close to that of the corotation drift.
Figure \ref{fig.driftscolat80} also shows that at $\delta$=10$^{\circ}$ the $\theta'$ component is dominant, and the largest contribution is from the curvature drift.

Particles that are propagating scatter-free (i.e.~for which  $\alpha$ quickly reaches values near zero) will be subject mainly to curvature drift. In the unipolar Parker field of Eqs.~(\ref{eq.br})--(\ref{eq.bphi}), or in the northern hemisphere of a bipolar $A^+$ configuration, near the heliographic equator, this drift has a $\theta'$ component that pushes positively charged ions downwards as they propagate, while its $\phi'$ component is close to zero (see Figure \ref{fig.fabsgvstheta}). At high heliographic latitudes both the $\theta'$ and $\phi'$ components will be significant. 

\begin{figure*}
\noindent\includegraphics[width=39pc]{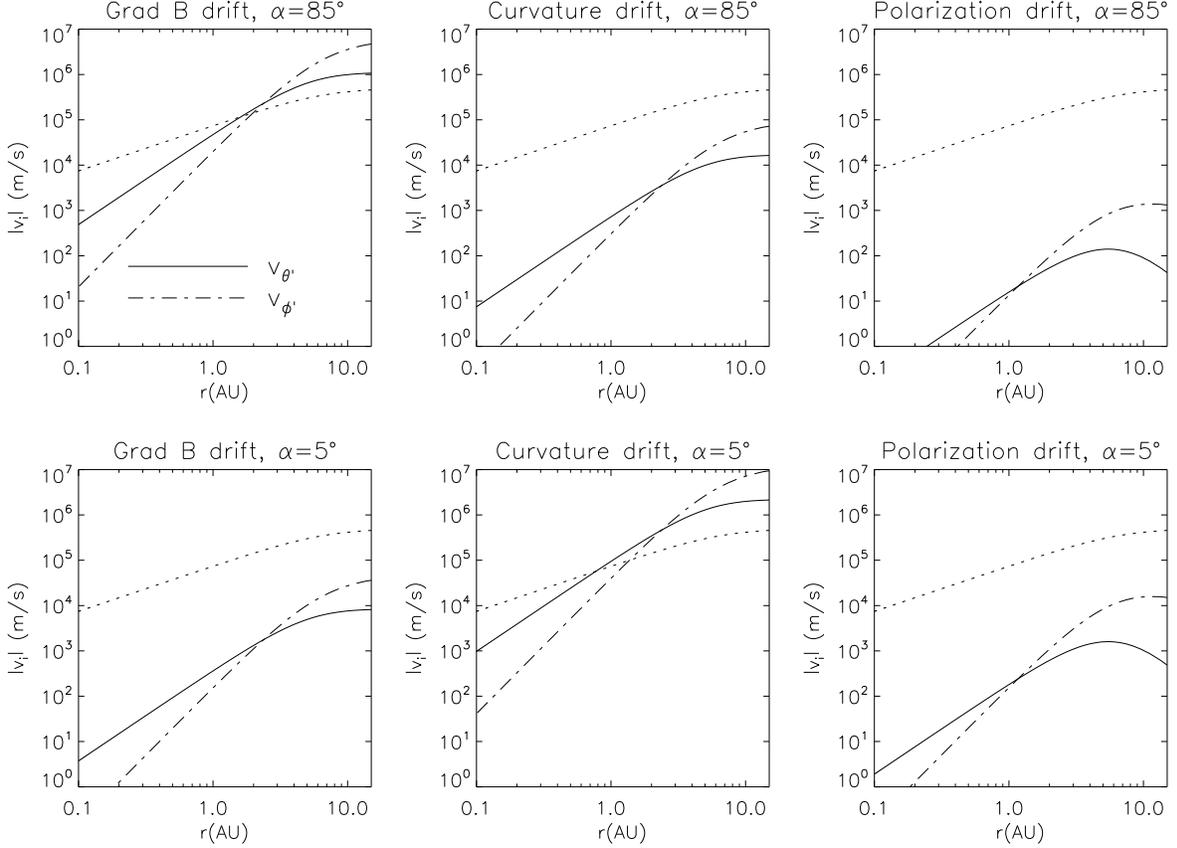}
\caption{Absolute values of drift velocity components for a 100 MeV proton at latitude $\delta$=80$^{\circ}$ as a function of $r$, in the same format as Fig.~\ref{fig.driftscolat80}.}
\label{fig.driftscolat10}
\end{figure*}

Figure \ref{fig.driftscolat10} shows the magnitude of drift velocity components for a particle at latitude $\delta$=80$^{\circ}$ ($\theta$=10$^{\circ}$). Here one can see that at this latitude and distances from the Sun greater than about 2 AU, drifts can be more than one order of magnitude larger than the corotation drift. Since at these locations the direction of the magnetic field is close to parallel to the Sun's rotation axis, drifts efficiently move particles approximately parallel to the heliographic equatorial plane.
At high heliolatitude and large $r$, the $\phi'$ component is the dominant drift.

Figure \ref{fig.driftscolat80} shows that drift velocity components tend to a constant value at large $r$. From Eqs.(\ref{eq.g_function})--(\ref{eq.f_function}) for $x$$\gg$1, we can see that the functions $g$ and $f$ tend to the following asymptotic values:
\begin{eqnarray}
g(r,\theta) &\sim& g_a(\theta)=  \frac{a}{B_0 r_0^2} \,  \cot{\theta}  \quad , \quad x \gg 1  \\
f(r,\theta) &\sim& f_a(\theta)=  \frac{a}{B_0 r_0^2}   \quad , \quad x \gg  1
\end{eqnarray}
which are constant at a given latitude.
Given that $x$ is directly proportional to $r \sin{\theta}$, at high latitudes a larger $r$ is required to ensure that the condition  $x$$\gg$1 is valid and the asymptotic values are reached.
Figures \ref{fig.driftscolat80} and \ref{fig.driftscolat10} show that close to the equatorial plane the asymptotic values are reached at distances of about 1 AU, while at high helioloatitudes this happens at distances beyond 10 AU.

Overall, Figures \ref{fig.driftscolat80} and \ref{fig.driftscolat10} show that drift velocities for 100 MeV protons are close to and in some cases much larger than the corotation drift velocity, and therefore should be taken into account as a possible source of cross field transport. 

It should be noted that BH1968 compared the magnitude of the curvature drift with that of the corotation drift, within their assumption of scatter-free propagation. It appears that there is a numerical error in their 
calculation of the ratio of these velocities (specifically in the numerical factor of the equation following their Eq.(20)). As a result, they incorrectly concluded that the curvature drift velocity is many orders of magnitude smaller than the corotation one, and consequently that the drift has a negligible $\theta$ component.

\subsection{Heavy ions}

The expressions for drift velocities derived in Section \ref{sec.unipolar} depend on the mass over charge ratio.
It is well known from measurements that SEP heavy ions are typically only partially ionised \citep{Kle2006}, making the value of $m_0/q$ large. It is therefore expected that for SEP heavy ions of energies of 100 MeV/nucleon, drift velocities will be larger than those of 100 MeV protons (shown in Figures \ref{fig.driftscolat80} and \ref{fig.driftscolat10}), by a factor $A/Q$ where $A$ is the ionic mass number and $Q$ the charge number.

\subsection{Evolution of SEP pitch-angle}

\begin{figure}
\begin{centering}
\noindent\includegraphics[width=15pc]{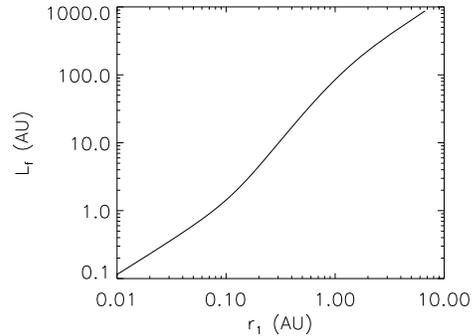}
\caption{Characteristic focussing radial distance $L_f$ versus the location $r_1$ at which the particle is injected with pitch-angle of 90$^{\circ}$.}
\label{fig.focussing_length}
\end{centering}
\end{figure}

It is well known that in the absence of scattering, particles injected at the Sun are rapidly focussed by the Parker spiral magnetic field, so that their pitch-angle quickly becomes close to zero.
If a particle finds itself with pitch-angle close to $\alpha$=90$^{\circ}$ (e.g. as a result of scattering) at a radial location $r_1$ in interplanetary space (e.g.~at 1 AU) and continues to propagate away from the Sun, it will experience focussing, however this will not be as effective as for a particle injected at the Sun.
We study this effect by defining a characteristic focussing radial distance $L_f(r_1)$ as the distance it takes for a particle to focus from $\alpha$=90$^{\circ}$ at $r_1$ to $\alpha$=5$^{\circ}$.
Figure \ref{fig.focussing_length} shows the dependence of  $L_f$ on the \lq injection\rq\ location $r_1$.
The plot shows that a particle starting with pitch-angle $\alpha$=90$^{\circ}$ near the Sun will be focused very quickly. However $L_f$ increases rapidly with radial distance of particle injection, and a particle having $\alpha$=90$^{\circ}$ at 1 AU (for example due to scattering) will be focused to a pitch-angle of 5$^{\circ}$ only by the time it has reached a radial distance of $\sim$80 AU. Consequently, if scattering is present in the interplanetary medium, a significant fraction of particles will be characterised by large and intermediate pitch-angles, making the grad-$B$ drift non-negligible.
Due to slow focussing away from the Sun, 
even a low level of scattering is sufficient to generate a pitch-angle distribution with significant population at pitch-angles near 90$^{\circ}$.

\section{Discussion and conclusions} \label{sec.discussion}

We analysed drifts in the Parker spiral IMF using single particle first-order guiding centre theory. The drift velocities first derived by BH1968 were calculated in a reference frame with an axis parallel to magnetic field and generalised to the case when scattering is present and to particles of relativistic energies. The resulting equations, Eqs.(\ref{eq.parkerelec1})--(\ref{eq.parkerelec3}) and Eqs.~(\ref{eq.vdl})--(\ref{eq.vdthetapr}), which include dependence on the particle's pitch angle, are applicable to SEPs and GCRs. 

We have shown that, contrary to current thinking, drift velocities are significant for protons at the upper end of the SEP energy range, and especially for heavy ions. 

Particles propagating scatter-free are subject to curvature drift, which near the heliographic equator is essentially a drift in latitude, and at high heliolatitudes includes both a $\phi'$ and a $\theta'$ component.
If scattering takes place in the IMF, particles can have a range of pitch-angles at large distances from the Sun, and both curvature and grad-$B$ drift can become significant. 

Drift velocities vary greatly with location in interplanetary space.
Not far above the heliographic equatorial plane, e.g.~at latitude $\delta$=10$^{\circ}$, the magnitudes of grad-$B$ and curvature drifts for 100 MeV protons can be as large as the corotation drift, reaching their maximum value at a distance of about 1 AU from the Sun. At high heliolatitudes, e.g.~$\delta$=80$^{\circ}$, magnitudes of grad-$B$ and curvature drifts greatly exceed the corotation drift at distances $>$2 AU and continue to grow with radial distance until $\sim$10 AU.

The large increase in drift velocities at high heliolatitudes is consistent with Ulysses observations of SEPs over the poles of the Sun, which showed that the heliolongitude of the solar event associated with the SEPs is not an important parameter in determining the characteristics of the event \citep{Dal2003a}.

For partially ionised heavy ion SEPs, drift velocities can become much larger than for protons of the same speed, due to the dependence on $A/Q$.

The overall amount of drift experienced by a particle is determined by how its position and pitch-angle vary over time as a result of propagation parallel and perpendicular to the field, focussing and scattering-induced pitch-angle changes.
Additional transport across the field will be caused by perpendicular propagation associated with turbulence in the IMF, which includes a contribution from field line random walk (see e.g.~\cite{Lai2013} and references therein).

Results of full orbit test particle simulations of SEP propagation in the Parker spiral are presented in a related paper \citep{Mar2013}. The simulations show significant particle drift in agreement with the analytical results obtained in this paper and allow the transport across the field to be quantified.

Drifts should therefore be taken into account in SEP models as a source of cross-field transport and the assumption that particles remain tied to field lines, in-built within many modelling and data analysis approaches, be revised.

\bibliographystyle{agufull08}

\bibliography{drifts_biblio}

\begin{thebibliography}{18}
\providecommand{\natexlab}[1]{#1}
\expandafter\ifx\csname urlstyle\endcsname\relax
  \providecommand{\doi}[1]{doi:\discretionary{}{}{}#1}\else
  \providecommand{\doi}{doi:\discretionary{}{}{}\begingroup
  \urlstyle{rm}\Url}\fi

\bibitem[{\textit{{Burger} and {Potgieter}}(1989)}]{Bur1989}
{Burger}, R.~A., and M.~S. {Potgieter} (1989), {The calculation of neutral
  sheet drift in two-dimensional cosmic-ray modulation models}, \textit{\apj},
  \textit{339}, 501--511, \doi{10.1086/167313}.

\bibitem[{\textit{{Burger} et~al.}(1985)\textit{{Burger}, {Moraal}, and
  {Webb}}}]{Bur1985}
{Burger}, R.~A., H.~{Moraal}, and G.~M. {Webb} (1985), {Drift theory of charged
  particles in electric and magnetic fields}, \textit{Astrophys.~Space Sci.},
  \textit{116}, 107--129, \doi{10.1007/BF00649278}.

\bibitem[{\textit{{Burns} and {Halpern}}(1968)}]{Bur1968}
{Burns}, J.~A., and G.~{Halpern} (1968), {Dynamics of a charged particle in a
  spiral field}, \textit{\jgr}, \textit{73}, 7377.

\bibitem[{\textit{{Dalla} et~al.}(2003)\textit{{Dalla}, {Balogh}, {Krucker},
  {Posner}, {M{\"u}ller-Mellin}, {Anglin}, {Hofer}, {Marsden}, {Sanderson},
  {Tranquille}, {Heber}, {Zhang}, and {McKibben}}}]{Dal2003a}
{Dalla}, S., A.~{Balogh}, S.~{Krucker}, A.~{Posner}, R.~{M{\"u}ller-Mellin},
  J.~D. {Anglin}, M.~Y. {Hofer}, R.~G. {Marsden}, T.~R. {Sanderson},
  C.~{Tranquille}, B.~{Heber}, M.~{Zhang}, and R.~B. {McKibben} (2003),
  {Properties of high heliolatitude solar energetic particle events and
  constraints on models of acceleration and propagation}, \textit{\grl},
  \textit{30}(19), 8035, doi:10.1029/2003GL017,139, \doi{10.1029/2003GL017139}.

\bibitem[{\textit{{Forsyth} et~al.}(1996)\textit{{Forsyth}, {Balogh}, {Smith},
  {Erd{\"o}s}, and {McComas}}}]{For1996}
{Forsyth}, R.~J., A.~{Balogh}, E.~J. {Smith}, G.~{Erd{\"o}s}, and D.~J.
  {McComas} (1996), {The underlying Parker spiral structure in the Ulysses
  magnetic field observations, 1990-1994}, \textit{\jgr}, \textit{101},
  395--404, \doi{10.1029/95JA02977}.

\bibitem[{\textit{{Jokipii} et~al.}(1977)\textit{{Jokipii}, {Levy}, and
  {Hubbard}}}]{Jok1977}
{Jokipii}, J.~R., E.~H. {Levy}, and W.~B. {Hubbard} (1977), {Effects of
  particle drift on cosmic-ray transport. I - General properties, application
  to solar modulation}, \textit{\apj}, \textit{213}, 861--868,
  \doi{10.1086/155218}.

\bibitem[{\textit{{Kelly} et~al.}(2012)\textit{{Kelly}, {Dalla}, and
  {Laitinen}}}]{Kel2012}
{Kelly}, J., S.~{Dalla}, and T.~{Laitinen} (2012), {Cross-field Transport of
  Solar Energetic Particles in a Large-scale Fluctuating Magnetic Field},
  \textit{\apj}, \textit{750}, 47, \doi{10.1088/0004-637X/750/1/47}.

\bibitem[{\textit{{Klecker} et~al.}(2006)\textit{{Klecker}, {M{\"o}bius}, and
  {Popecki}}}]{Kle2006}
{Klecker}, B., E.~{M{\"o}bius}, and M.~A. {Popecki} (2006), {Solar Energetic
  Particle Charge States: An Overview}, \textit{Space Sci.~Revs.},
  \textit{124}, 289--301, \doi{10.1007/s11214-006-9111-0}.

\bibitem[{\textit{{Laitinen} et~al.}(2013)\textit{{Laitinen}, {Dalla}, and
  {Marsh}}}]{Lai2013}
{Laitinen}, T., S.~{Dalla}, and M.~S. {Marsh} (2013), {Energetic particle
  cross-field propagation early in a solar event.}, \textit{\apj},
  \textit{submitted,}.

\bibitem[{\textit{{Lario} et~al.}(1998)\textit{{Lario}, {Sanahuja}, and
  {Heras}}}]{Lar1998}
{Lario}, D., B.~{Sanahuja}, and A.~M. {Heras} (1998), {Energetic Particle
  Events: Efficiency of Interplanetary Shocks as 50 keV $<$ E $<$ 100 MeV
  Proton Accelerators}, \textit{\apj}, \textit{509}, 415--434,
  \doi{10.1086/306461}.

\bibitem[{\textit{{Luhmann} et~al.}(2007)\textit{{Luhmann}, {Ledvina},
  {Krauss-Varban}, {Odstrcil}, and {Riley}}}]{Luh2007}
{Luhmann}, J.~G., S.~A. {Ledvina}, D.~{Krauss-Varban}, D.~{Odstrcil}, and
  P.~{Riley} (2007), {A heliospheric simulation-based approach to SEP source
  and transport modeling}, \textit{Advances in Space Research}, \textit{40},
  295--303, \doi{10.1016/j.asr.2007.03.089}.

\bibitem[{\textit{{Marsh} et~al.}(2013)\textit{{Marsh}, {Dalla}, {Kelly}, and
  {Laitinen}}}]{Mar2013} 
{Marsh}, M.~S., S.~{Dalla}, J.~{Kelly}, and T.~{Laitinen} (2013), {Drift
  induced perpendicular transport of Solar Energetic Particles.},
  \textit{\apj}, \textit{in press}, ArXiv e-prints, arXiv:1307.1585

\bibitem[{\textit{{Northrop}}(1963)}]{Nor1963}
{Northrop}, T. (1963), \textit{{The adiabatic motion of charged particles.}},
  {Interscience}, {New York}.

\bibitem[{\textit{{Parker}}(1958)}]{Par1958}
{Parker}, E.~N. (1958), {Dynamics of the Interplanetary Gas and Magnetic
  Fields.}, \textit{\apj}, \textit{128}, 664, \doi{10.1086/146579}.

\bibitem[{\textit{{Roelof}}(1969)}]{Roe1969}
{Roelof}, E.~C. (1969), {Propagation of Solar Cosmic Rays in the Interplanetary
  Magnetic Field}, in \textit{Lectures in High-Energy Astrophysics}, edited by
  H.~{{\"O}gelman} and J.~R. {Wayland}, p. 111.

\bibitem[{\textit{{Rossi} and {Olbert}}(1970)}]{Ros1970}
{Rossi}, B., and S.~{Olbert} (1970), \textit{{Introduction to the physics of
  space.}}, {McGraw-Hill}, {New York}.

\bibitem[{\textit{{Ruffolo}}(1995)}]{Ruf1995}
{Ruffolo}, D. (1995), {Effect of adiabatic deceleration on the focused
  transport of solar cosmic rays}, \textit{\apj}, \textit{442}, 861--874,
  \doi{10.1086/175489}.

\bibitem[{\textit{{Winge} and {Coleman}}(1968)}]{Win1968}
{Winge}, C.~R.~J., and P.~J.~J. {Coleman} (1968), {The motion of charged
  particles in a spiral field}, \textit{\jgr}, \textit{73}, 165.

\end{thebibliography}


%
%
%
%
%
%
%

\begin{acknowledgments}
This work has received funding from the European Union Seventh
Framework Programme (FP7/2007-2013) under grant agreement n.~263252
[COMESEP]. TL acknowledges support from the UK Science and Technology
Facilities Council (STFC) (grant ST/J001341/1).
\end{acknowledgments}

\end{article}


%
%

%
%
%
%
%
%
%


\end{document}